# An Algorithm for Detection of Selfish Nodes in Wireless Mesh Networks

Jaydip Sen[1], and Kaustav Goswami[2]
Innovation Lab, Tata Consultancy Services Ltd.
Bengal Intelligent Park, Salt Lake Electronics Complex, Kolkata, India
Emails: {[1]Jaydip.Sen, [2]Kaustav.G}@tcs.com

*Abstract*—Wireless mesh networks (WMNs) are evolving as a key technology for next-generation wireless networks showing raid progress and numerous applications. These networks have the potential to provide robust and high-throughput data delivery to wireless users. In a WMN, high speed routers equipped with advanced antennas, communicate with each other in a multi-hop fashion over wireless channels and form a broadband backhaul. However, the throughput of a WMN may be severely degraded due to presence of some selfish routers that avoid forwarding packets for other nodes even as they send their own traffic through the network. This paper presents an algorithm for detection of selfish nodes in a WMN. It uses statistical theory of inference for reliable clustering of the nodes and is based on local observations by the nodes. Simulation results show that the algorithm has a high detection rate while having a low rate of false positives.

## I. INTRODUCTION

Wireless mesh networking has emerged as a promising concept to meet the challenges in next-generation networks such as providing flexible, adaptive, and reconfigurable architecture while offering cost-effective solutions to the service providers [1]. Unlike traditional Wi-Fi networks, with each access point (AP) connected to the wired network, in WMNs only a subset of the APs are required to be connected to the wired network. The APs that are connected to the wired network are called the Internet gateways (IGWs), while the APs that do not have wired connections are called the mesh routers (MRs). The MRs are connected to the IGWs using multi-hop communication. Due to the recent research advances in WMNs, these networks have been used in numerous applications such as in home networking, community and neighborhood monitoring, security surveillance systems, disaster management and rescue operations etc [2].

In a community-based WMN, a group of MRs managed by different operators form an access network to provide last-mile connectivity to the Internet. As with any end-user supported infrastructure, ubiquitous cooperative behavior in these networks cannot be assumed a priori. Preserving scarce access bandwidth and power, as well as security concerns may induce some selfish users to avoid forwarding data for other nodes, even as they send their own traffic through the network. However, the selfish behavior of an MR degrades the performance of a WMN since it increases the latency in packet delivery and packet drops and decreases the network throughput.

To enforce cooperation among nodes and detect selfish nodes in ad hoc wireless networks, various collaboration schemes have been proposed in the literature [3]. Majority of these proposals are based on trust and reputation frameworks which attempts to identify misbehaving nodes by suitable decision making systems and then isolate or punish them. The reputation of participating nodes is built based on local observation at the node, second-hand observation at other nodes or both.

To address the issue of selfish nodes in a WMN, this paper presents a scheme that uses local observations in the nodes for detecting node misbehavior. The scheme is applicable for on-demand routing protocol like AODV, and uses statistical theory of inference and clustering techniques to make a robust and reliable classification (cooperative or selfish) of the nodes based on their packet forwarding activities as observed by their neighbors. In addition, it introduces some additional fields in the packet header for AODV protocol so that detection accuracy is increased.

The rest of the paper is organized as follows. Section II presents some related work. Section III gives a brief background of the AODV protocol and a finite state machine model of the local observations of a node. The proposed scheme is described in Section III. Section IV presents simulations results, and Section V concludes the paper while identifying some potential future work.

## II. RELATED WORK

The concept of neighborhood monitoring to check the activities of other nodes has been proposed by researchers in a number of mechanisms especially in the context of wireless ad hoc networks. The idea of watchdog mechanism to monitor neighbors was first proposed by Marti et al [4]. A scheme named *pathrater* was also proposed to avoid misbehaving nodes in routing. Buchegger and Boudec have proposed the CONFIDANT protocol that assigns a rating for every node in an ad hoc network based on watchdog and second-hand rating information gathered from other nodes [5].

Mahajan et al have proposed a mechanism named CATCH [6], which consists of two modules: (i) *anonymous challenge message* (ACM) and (ii) *anonymous neighbor verification* (ANV). First, an ACM message from an unknown sender is sent to all its neighbors. As the sender is unknown, all the nodes further broadcast the ACM message. In the ANV phase, a tester node sends cryptographic hash of a random token for rebroadcast and also records other hashes sent by others. The tester node releases the secret token to another node which successfully authenticates itself.

Vigna et al have proposed an approach to detect intrusions in AODV that works by stateful signature-based analysis of the observed traffic [7]. Sensors are placed on selected nodes for promiscuous sensing of radio channels. Each sensor has database of attack signatures and looks for a signature match in the traffic. A match triggers a response, usually an alert.

Pirzada and McDonald [8] have described a model of building trust relationship between nodes in an ad hoc network. The nodes passively monitor the packets received and forwarded by other nodes and compute the trust values for their neighbors. The trust values are used for computing the trustworthiness of links. For routing, links with high trust values are chosen so as to avoid the malicious and selfish nodes.

Conti et al [9] have proposed a scheme in which a node exploits its local knowledge to estimate the reliability of a path. Unlike the conventional method of denying selfish users, it provides a degraded service to these nodes by selective slow packet forwarding.

Santhanam et al [10] have presented a mechanism to judge a node's behavior based on observed traffic reports submitted to local sink agents, dispersed throughout the network. The sink nodes apply a set of forwarding rules to isolate a selfish node based on the number of times it is caught in selfish acts. The scheme is independent of the routing protocol or network architecture, and is suitable for multi-channel wireless mesh network.

Tseng et al. have applied techniques based on finite state machines to detect misbehaving nodes in AODV routing protocol [11]. The approach involves monitoring nodes that cooperate with each other and aggregate their observations at different locations in the network.

Yang et al [12] have described the SCAN protocol that addresses two issues simultaneously: (i) routing (control packets) misbehavior, and (ii) forwarding (data packets) misbehavior. Each node monitors its neighbors independently and the nodes in a neighborhood collaborate with each other through a distributed consensus protocol.

The proposed mechanism in this paper relies on local observation of each node in a WMN. Based on the local information in each node and using a finite state machine model of the AODV protocol, a robust statistical theory of estimation is applied to identify selfish nodes in the network. The proposed mechanism is a modification of the protocol proposed in [13]. The objective of the proposed mechanism is to achieve higher detection efficiency by exploiting the information in some additional fields in the packet header in AODV routing. This is explained in Section IV.

III. AODV AND MODELING OF THE STATE MACHINE

*Ad hoc on-demand distance vector* (AODV) routing protocol uses an on-demand approach for finding routes to a destination node. It employs destination sequence numbers to identify the most recent path. The source node and the intermediate nodes store the next-hop information corresponding to each flow for data packet transmission. The source node floods the *route request* (RREQ) packet in the network when a route is not available for the desired destination. It may obtain multiple routes to different destinations from a single RREQ. The RREQ carries the source identifier (*src_id*), the destination identifier (*dest_id*), the source sequence number (*src_seq_num*), the destination sequence number (*dest_seq_num*), the broadcast identifier (*bcast_id*), and the time to live (TTL). When an intermediate node receives a RREQ, it either forwards the request further or prepares a *route reply* (RREP) if it has a valid route to the destination. Every intermediate node, while forwarding a RREQ, enters the previous node address and its BcastID. A timer is used to delete this entry in case a RREP is not received before the timer expires. This helps in storing an active path at the intermediate node as AODV does not employ source routing of data packets. When a node receives a RREP packet, information of the previous node from which the packet was received is also stored, so that data packets may be routed to that node as the next hop toward the destination.

It is clear that AODV depends heavily on cooperation among the nodes for its successful operation. A selfish node can easily manipulate the protocol to minimize its chances of being included on routes for it is neither the source nor the destination. It may drop or tamper with the RREQ messages to ensure that no routes will ever be selected through it. Alternatively, it may drop, delay, or modify the RREP messages so as to prevent the replies from reaching the source node. The proposed mechanism attempts to detect selfish nodes in a WMN so that these nodes may be isolated from the network. In the following subsection, the finite state machine model of the protocol is presented.

*A. Finite State Machine Model*

In the proposed mechanism, with AODV as the underlying routing protocol, the set of all messages corresponding to a RREQ flooding and the unicast RREP is referred to as a *message unit*. It is clear that no node in the network can observe all the transmission in a message unit. The subset of a message unit that a node can observe is referred to as the *local message unit* (LMU). The LMU for a particular node consists of the messages transmitted by that node, the messages transmitted by all its neighbors, and messages overheard by the node. The detection of selfish nodes is made on the basis of data collected by each node from its observed LMUs.

Corresponding to each message transmission in an LMU, a node maintains a record of its sender, and the receiver in its neighborhood. It also keeps record of the neighbor nodes that receive the RREQ broadcast messages sent by the node itself. The messages are assumed to follow the sequence of the AODV protocol.

The finite state machine shown in Fig. 1 depicts various states through which a neighbor node undergoes for each LMU [13]. The corresponding states for the numbers mentioned in Fig.1 can be found in Table I. To distinguish the finals states, these states are shaded. Every message transmission by a node causes a state transition in each of its neighbor's finite state machine. The finite state machine in one neighbor node gives only a local view of the activities of the node being monitored. It does not in any way, represents the actual behavior of the monitored node. The collaborative participation of each neighbor node makes it possible to get an accurate global picture regarding the monitored node's behavior. In the rest of the paper, a node whose activity is being monitored by its neighbors is referred to as a *monitored node*, and its neighbors are referred to as a *monitor node*. In the proposed protocol, each node plays the dual role of a monitor node and a monitored node for each of its neighbors.

TABLE I. STATES OF FINITE STATE MACHINE FOR AN LMU

| State | Interpretation |
|---|---|
| 1: init | Initial phase; no RREQ is observed |
| 2: unexp RREP | Receipt of a RREP without RREQ observed |
| 3: rcvd RREQ | Receipt of a RREQ observed |
| 4: fwd RREQ | Broadcast of a RREQ observed |
| 5: timeout RREQ | Timeout after receipt of RREQ |
| 6: rcvd RREP | Receipt of a RREP observed |
| 7: LMU complete | Forwarding of a valid a RREP observed |
| 8: timeout RREP | Timeout after receipt of a RREP |

Each monitor node in the network observes a series of interleaved LMUs for a routing session. Each LMU can be identified by the source-destination pair contained in a RREQ message. Let us denote the $k^{th}$ LMU observed by a monitor node as $(s_k, d_k)$. The pair $(s_k, d_k)$ does not uniquely identify a LMU as a source can issue multiple RREQs for the same destination. However, since the subsequent RREQs have some delays associated with them, we can safely assume that there is only one active LMU $(s_k, d_k)$ in the network at any point of time.

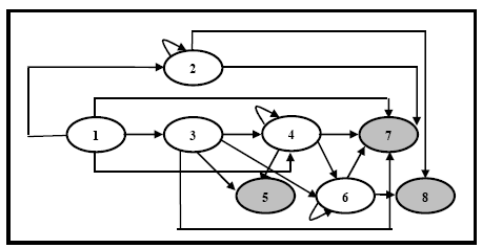

Figure 1. Finite state machine of a monitored node

Before the beginning of a routing process, a monitored node starts with the state 1 in its finite state machine. As the monitor node(s) observes the behavior of the monitored node by examining the LMUs, it records a sequence of transitions form its initial state 1 to one of its possible final states - 5, 7 and 8.

When a monitor node broadcasts a RREQ, it assumes that the monitored node has received it. The monitor node, therefore, records a state transition 1 → 3 for the monitored node's finite state machine. If a monitor node observes a monitored node to broadcast a RREQ, then a state transition of 3 → 4 is recorded if the RREQ message was previously sent by the monitor node to the monitored node; otherwise a transition of 1 → 4 will be recorded meaning thereby that the RREQ was received by the monitored node from some other neighbor. The transition to a timeout state occurs when a monitor node finds no activity by the monitored node for the concerned LMU before the expiry of a timer. When a monitor node observe a monitored node to forward a RREP, it records a transition to the final state – *LMU complete* (state no 7). At this state, the monitored node becomes a candidate for inclusion on a routing path.

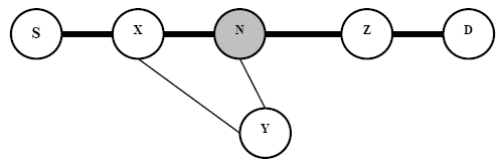

Figure 2. An example LMU observed by node *N*

Fig. 2 depicts an example of LMU observed by the node *N* during the discovery of a route from the source node *S* to the destination node *D* indicated by bold lines. Table II shows the events observed by node *N* and the corresponding state transitions for each of its three neighbor nodes *X*, *Y* and *Z*.

When the final state is reached, the finite state machine terminates and the corresponding sequences of state transitions are stored by each node for each of its neighbors. When sufficient number of events is collected by a node, a statistical analysis is performed to detect the presence of any selfish nodes in the network.

TABLE II. STATE TRANSITIONS OF THE NEIGHBORS OF NODE N

| Neighbor | Events | State changes |
|---|---|---|
| X | X broadcasts RREQ | 1 → 4 |
|   | N broadcasts RREQ | 4 → 4 |
|   | N sends RREP to X | 4 → 6 |
|   | X sends RREP to S (overheard) | 6 → 7 |
| Y | Y broadcasts RREQ | 1 → 4 |
|   | N broadcasts RREQ | 4 → 4 |
|   | Timeout | 4 → 5 |
| Z | N broadcasts RREQ | 1 → 3 |
|   | Z broadcasts RREQ | 3 → 4 |
|   | Z sends RREP to N | 4 → 7 |

IV. THE PROPOSED ALGORITHM

As mentioned in the previous section, a monitoring node keeps a record of state transitions in the finite state machine of a monitored node in each LMU. These sequences can be represented as a transition matrix $T = [T_{ij}]$, where $T_{ij}$ is the number of times the transition $i \to j$ is found. The monitor node invokes a detection algorithm every $W$ seconds using data from the most recent $D = d * W$ seconds of observations, where $d$ is a small integer. The parameter $D$, called the *detection window*, should be such that it is possible to punish the selfish nodes promptly while maintaining a high level of accuracy.

In the following subsection, some of the important issues in designing the proposed algorithm are discussed.

*A. Issues in Algorithm Design*

While a transition matrix summarizes the local routing behavior of a monitored node, it is not possible to determine selfish behavior of a node based on its local transition probabilities only. By comparing the transition matrices of a collection of nodes, one might be able to detect selfish nodes with higher confidence than looking at the transition matrix of each node individually. Keeping this in mind, the proposed algorithm initially clusters the neighbors of a monitoring node and then classifies the clusters into selfish or cooperative nodes. Some of the issues in forming the clusters are as follows:

First, to make the clustering algorithm robust in presence of noise in the data, a statistical theory of inference-based approach is followed which takes into account the pair-wise comparisons of the transition matrices of each pair of nodes. Second, to reliably identify which cluster contains the selfish nodes, an additional measure of cooperation, called *cooperation index*, for the nodes. The cluster which has its cooperation index less than a threshold value is assumed to contain the selfish nodes. Finally, a test is developed based on the *analysis of variance* (ANOVA) among the clusters to determine whether clustering is informative to the purpose of classification.

The proposed algorithm is described in detail in the following subsection.

*B. The Detection Algorithm*

In the proposed algorithm, a node is assumed to monitor the activities of its $R$ neighbors which are identified by their respective indices $1, 2,….R$. Let $T^{(r)} = [f_{ij}^{(r)}]$ denote the observed transition matrix for the $r^{th}$ neighbor, where $[f_{ij}^{(r)}]$ is the number of transitions from state $i$ to state $j$ observed in the previous detection window. If $m$ is the number of states in the finite state machine in each node, the size of $T^{(r)}$ is $m$ x $m$. Let $T_i^{(r)} = [f_{i1}^{(r)},...f_{im}^{(r)}]$ denote the $i^{th}$ row of the transition matrix $T^{(r)}$, which shows the transitions out of state $i$ at the neighbor node $r$. If two neighbor nodes $r$ and $s$ have identical distributions corresponding to transitions from state $i$, then one can write $T_i^{(r)} \equiv T_i^{(s)}$.

To test the hypothesis $T_i^{(r)} \equiv T_i^{(s)}$ the Pearson's $\chi^2$ test is used as follows.

$$\chi^2(i) = \frac{\sum_{l\varepsilon(r,s)} \sum_{j=1}^{m} [f_{ij}^{(l)} - \bar{f}_{ij}^{(l)}]^2}{\bar{f}_{ij}^{(l)}} \quad (1)$$

$$\bar{f}_{ij}^{(l)} = F_{ij}^{(l)} \frac{f_{ij}^{(r)} + f_{ij}^{(s)}}{F_i^{(r)} + F_i^{(s)}}$$

where, $F_i^{(r)}$ and $F_i^{(s)}$ denote total number of transitions for state I in $T^{(r)}$ and $T^{(s)}$ respectively.

If the value of $\chi^2$ exceeds the value of $\chi^2_{m-1,\alpha}$ then the hypothesis $T_i^{(r)} \equiv T_i^{(s)}$ is rejected at confidence interval $\alpha$. If we write $K_i^{rs}$ for the event that $\chi^2_{(i)} > \chi^2_{m-1,\alpha}$, then the conditional probability $P(T_i^{(r)} \equiv T_i^{(s)} | B_i^{rs})$ can be taken as a reasonable estimator of the similarity between $r$ and $s$ with respect to the state $i$. In absence of any prior information, it is reasonable to assume that $r$ and $s$ have no similarity in state $i$ and the probability that the Pearson test rejects its hypothesis to be 0.5 [13].

In order to evaluate the similarity between $r$ and $s$ for all the $m$ states, (1) is applied to all rows of $T^{(r)}$ and $T^{(s)}$. This yields a vector $B^{(rs)} = [B_i^{(rs)}]$, $\{i = 1,…m\}$. From the standard Markovian principle one can write:

$$L_{rs} = P(T^{(r)} \equiv T^{(s)} | B^{(rs)})$$

$$= \alpha^{S^{(rs)}}(1-\alpha)^{m-S^{(rs)}}$$

$$\approx \alpha^{S^{(rs)}} \quad (2)$$

where $\quad S^{(rs)} = \sum_{i=1}^{m} B_i^{(rs)} \quad (3)$

the lower-order terms in the right hand side of (3) are ignored since α << 1.

For small value of α, $L_{rs}$ monotonically decreases in $S^{(rs)}$, which, as evident from (3), is the number of rejections of Pearson's hypothesis. Therefore, 1 -- $L_{rs}$ may be taken as the measure of the dissimilarity between the neighbor nodes $r$ and $s$.

In presence of noise in the data, however, it is found that for two nodes $r$ and $s$ which have $L_{rs} \approx 1$, a third node $t$ may cause inconsistency such that $L_{rt} \neq L_{st}$. To avoid this inconsistency in clustering in the proposed algorithm, clustering are not computed on the basis of pair-wise dissimilarity. To compute dissimilarity between $r$ and $s$, the $L$ values for all neighbors are computed with respect to $r$ and $s$ separately, and the following equation is applied:

$$d_{rs} = 1 - \frac{n_{rs}^2}{n_{r/s} * n_{s/r}} \quad (4)$$

where

$$n_{rs} = \sum_{t \neq r,s} \min(L_{rt}, L_{st}),$$

$$n_{r/s} = \sum_{t \neq r,s}^{K} L_{rt}$$

$$n_{s/r} = \sum_{t \neq r,s}^{K} L_{st}.$$

It may be observed that the computation of $d_{rs}$ does not involve on $L_{rs}$ - the pair-wise similarity between nodes $r$ and $s$. In fact, it measures the degree of inconsistency in similarity between $r$ and $s$ with all their neighbors. Since, in the computation, contribution of each neighbor plays its role, $d_{rs}$ presents a robust indicator for dissimilarity between nodes and plays a crucial part in computing the clusters [13]. For clustering, an *agglomerative hierarchical clustering* technique is used. This is a single-linkage approach in which each cluster is represented by all of the objects in the cluster, and the similarity between two clusters is measured by the similarity of the closest pair of data points belonging to different clusters. The cluster merging process repeats until all the objects are eventually merged to form one cluster [14].

After the nodes are clustered into similar sets, the sets are further classified into three groups: (i) a set ($G$) of cooperative nodes, (ii) a set ($B$) of selfish nodes, and (iii) a set of nodes whose behavior could not be ascertained. The cooperation score ($C_r$) of a node is computed as [13]:

$$C_r = \frac{\sum_{i,j \varepsilon G}^{m} n_{ij}^{(r)}}{|G|} - \frac{\sum_{i,j \varepsilon B}^{m} n_{ij}^{(r)}}{|B|} \qquad (5)$$

The set $B$ is most likely to contain the selfish nodes. To reduce false positives (i.e. wrongly identifying a cooperative node as selfish), an ANOVA test is applied as described in [13]. The ANOVA approach computes a probability $P_k$ of the random variation among the mean cooperation scores of $k$ clusters. A lower value of $P_k$ implies that the clusters actually represent distinct differences in their behavior. At each iteration, $k$ clusters are formed and $P_k$ is compared with a pre-defined level of significance β. If $P_k$ < β, clusters are believed to be reliably reflecting the behavior of the nodes and their classifications are accepted. The cluster with lowest mean cooperation score is assumed to contain the selfish nodes. If $P_k$ > $P_{k-1}$, the neighbor behavior has not been properly reflected in the cluster formation, which has led to the increase in the value of $P_k$. In this case, all the nodes are classified as cooperative, and the next iteration of the algorithm is executed. The confidence parameter β can be tuned so as to adjust the alacrity of detection of selfish nodes and rate of false positives [13].

In spite of all the above statistical approaches, there is still a possibility of misclassification. The proposed algorithm further reduces the probability of misclassification by a new cross-checking mechanism. For this purpose, a minor modification is suggested in the packet header for AODV routing. Two additional fields are inserted in the header of a RREQ packet. These fields are: *next_to_source* and *duplicate_flag* to indicate respectively the address of the node that is next hop to the source, and whether the packet is a duplicate packet which has already been broadcasted by some other nodes in the network. In the header of a RREQ packet, in addition to the above two fields, another field called *next_to_destination* is added to indicate the address of the node to which the packet must be forwarded in the reverse path. It has been shown in [16], with the above extra fields, it is possible to detect every instance of selfish behavior in a wireless network with 100% detection accuracy, if the following conditions are satisfied: (i) no packet loss lost due to interference, (ii) links are bi-directional, (iii) the nodes are stationary, and (iv) the queuing delays are bounded. Since all these conditions cannot be guaranteed in a real-world deployment, there will be always some detection inaccuracy. However, the combined approach of the robust clustering and the monitoring of packets with additional fields substantially increases the detection efficiency and reduces the false positives as will be evident from the simulation results.

V. SIMULATION RESULTS

The proposed protocol is evaluated with network simulator *ns-2* (version 2.29) [15] with parameters presented in Table III. The objective is to evaluate the efficiency of the algorithm and compare its performance with the protocol proposed by Wang et al in [13].

At the start of the simulation, a fraction of nodes are chosen randomly as the selfish nodes. A selfish node adopts either of the two strategies--dropping RREQs (DROP_REQ) or dropping RREPs (DROP_REP). In both cases, control packets are dropped with a constant probability. For DROP_REP, a selfish node always rebroadcasts RREQs even if it has a route in its cache. To evaluate the detection efficiency and speed, the packet dropping probability is varied from 1.0 to 0.1. β is chosen as 0.4 to have the best tradeoff between detection rate and false positive rate.

TABLE III. SIMULATION PARAMETERS

| Parameter | Value |
| --- | --- |
| Simulation area | 900 m * 900 m |
| Simulation duration | 1600 sec |
| No. of nodes in the network | 50 |
| MAC protocol | 802.11b |
| Routing protocol | AODV |
| Raw channel bandwidth | 11 Mbps |
| Traffic type | CBR UDP |
| Network traffic volume | 60 packets / sec |
| Packet size | 512 bytes |
| Time-out for RREQ broadcast | 0.5 sec |
| Time-out for receiving RREP | 3 sec |
| Pearson confidence (α) | 0.1 |
| Observation window ($W$) | 100 sec |
| Detection window ($D$) | 400 sec |
| Session arrival distribution | Poisson |
| Session duration distribution | Exponential |

Fig 3 and Fig 4 represent respectively the detection rate and the false alarm rate with 50% nodes in the network configured as selfish and dropping RREQs (i.e. DROP_REQ). The results are the average of 10 runs of the simulation. The proposed algorithm performs better than Wang's algorithm since it doubly checks the detection results- one from the clustering and the other from the routing header information to make a more reliable detection.

Fig. 5 and Fig. 6 show that the packet dropping (DROP_REP) has no impact on the detection rate and the false positive rate when 50% nodes in the network are acting as selfish nodes. This difference in behavior in case of DROP_REQ and DROP_REP lies in the fact that while RREQ is a broadcast message sent from the source, the RREP is sent in a single path by the destination in a unicast manner. Since RREP involves very few number of nodes, for majority of the nodes the state machine will terminate in state 5, instead of states 7 and 8. It is evident from Fig 5. and Fig. 6 that the proposed algorithm has an average 80% increase in detection rate and 50 % reduction in false positives.

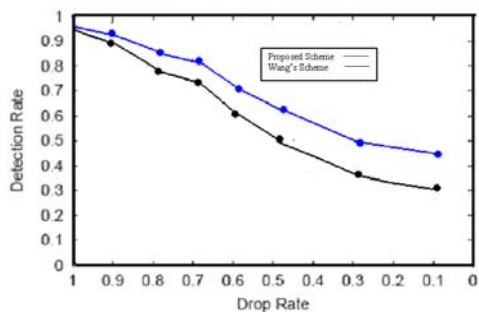

Figure 3. Detection rate in DROP_REQ scenario

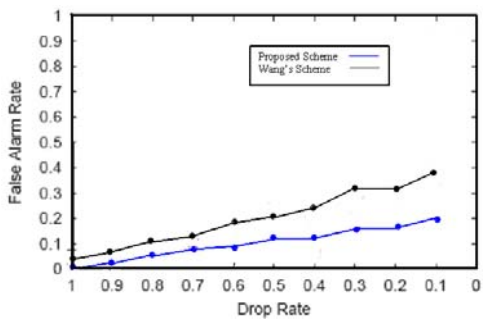

Figure 4. Detection rate in DROP_REQ scenario

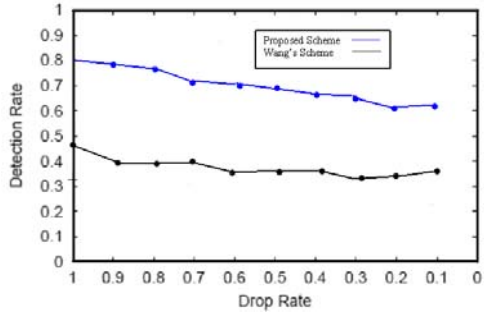

Figure 5. Detection rate in DROP_REP scenario

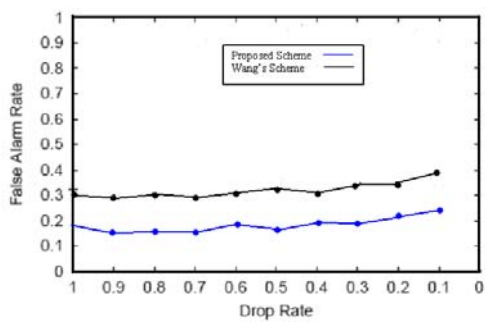

Figure 6. False alarm rate in DROP_RREP scenario

## VI. CONCLUSION AND FUTURE WORK

Detection of selfish nodes is crucial in WMNs since these nodes don't forward packets for other nodes and degrade the performance of the networks. This paper has presented a statistical theory of inference-based clustering algorithm for detection of selfish nodes. The routing protocol is assumed to be AODV and a finite state machine model is developed based on locally observed messages in each node. To increase the reliability of clustering, an ANOVA test is applied and finally, a new cross-checking mechanism is used by inserting extra fields in the packet headers. Simulation results show that the algorithm has better detection efficiency and reduced false alarm rates compared to some existing mechanisms.

REFERENCES

[1] I.F. Akyildiz, X. Wang, and W. Wang, "Wireless mesh networks: a survey," *Computer Networks*, Vol 47, No 4, March 2005, pp. 445- 487.
[2] A.A. Franklin and C. S.R Murthy, "An introduction to wireless mesh networks," in *Security in Wireless Mesh Networks*," Y. Zhang, J. Zheng, and H. Hu, Eds. CRC Press, 2007, pp. 3-44.
[3] L. Santhanam, B. Xie, and D.P. Agrawal, "Selfishness in mesh networks: wired multihop MANETs," *IEEE Wireless Comm. Magazine*, Vol 15, No 4, August 2008, pp. 16-23.
[4] S. Marti, T.J. Giuli, K. Lai, and M. Baker, "Mitigating routing misbehavior in mobile ad hoc networks," in *Proceeedings of MobiCom 2000*, pp. 255-265.
[5] S. Buchegger and J.-Y. L. Boudec, "Performance analysis of the CONFIDANT protocol: Cooperation of nodes-fairness in dynamic ad-hoc networks," In *Proc. of MobiHoc*, 2002, pp. 226-236.
[6] R. Mahajan, M. Rodrig, D. Wetherall, and John Zahorjan , "Sustaining cooperation in multihop wireless networks," in *Proc. of NSDI 2005*, Vol 2, pp. 231-244.
[7] G. Vigna, S. Gwalani, K. Srinivasan, E.M. Belding-Royer, and R.A. Kemmerer, "An intrusion detection tool for AODV-based ad hoc wireless networks," in *Proc of Annual Comp. Sec. Appl. Conf (ACSAC) 2004*, pp. 16-27.
[8] A. Pirzada and C. McDonald, "Establishing trust in pure ad hoc networks," in *Proceedings of the 27$^{th}$ Australian Conference on Computer Science, 2004*, pp. 181-199.
[9] M. Conti, E. Gregori, and G. Maselli, "Reliable and efficient forwarding in MANETs," *Ad Hoc Networks Journal*, Vol 4, No 3, 2006, pp. 398-415.
[10] L. Santhanam et al, "Distributed self-policing architecture for fostering node cooperation in wireless mesh networks," in *Proc. of PWC 2006*, Vol 4217, pp. 147-158.
[11] C.Y. Tseng, P. Balasubramanyam, C. Ko, R. Limprasittiporn, J. Rowe, and K. Levitt, "A specification-based intrusion detection system for AODV," in *Proc. of the 1$^{st}$ ACM Workshop on Security of Ad Hoc and Sensor Networks, 2003*, pp. 125-134.
[12] H. Yang, J. Shu, X. Meng, and S. Lu, "SCAN: self-organized network-layer security in mobile ad hoc networks," *IEEE Journal on Selected Areas in Communications*, Vol 24, 2006, pp. 261-273.
[13] B. Wang, S. Soltani, J.K. Shaprio, P-N. Tan, and M. Mutka, "Distributed detection of selfish routing in wireless mesh networks," Technical Report – MSU-CSE-06-19, Depart of Comp. Sc. and Engg, Michigan State University.
[14] W.F. Eddy, A. Mockus, and S. Oue, "Approximate single linkage cluster analysis of large datasets in high dimensional spaces," *Computational Statistics and Data Analysis*, Vol 23, pp. 29-43, 1996.
[15] Network simulator. URL: http://www.isi.edu/nsnam/ns
[16] H.J. Kim and J.M. Peha, "Detecting selfish behavior in a cooperative commons," In *Proc. of IEEE DySPAN, 2008*, pp. 1-12.